\documentstyle[12pt]{article}
\begin{document}
\newcommand{\be}{\begin{equation}}
\newcommand{\ben}{\begin{subequations}}
\newcommand{\een}{\end{subequations}}
\newcommand{\beq}{\begin{eqalignno}}
\newcommand{\eeq}{\end{eqalignno}}
\newcommand{\ee}{\end{equation}}
\newcommand{\wt}{\widetilde}
\def\lsim{\:\raisebox{-0.5ex}{$\stackrel{\textstyle<}{\sim}$}\:}
\def\gsim{\:\raisebox{-0.5ex}{$\stackrel{\textstyle>}{\sim}$}\:}
\renewcommand{\thefootnote}{\fnsymbol{footnote}}

\begin{center}
{\Large \bf  Some Comments on ``Split'' Supersymmetry} \\
\vspace{10mm}
Manuel Drees \\
\vspace{5mm}
{\it Physikalisches Institut der Universit\"at Bonn, Nussallee 12, 53115 Bonn,
  Germany} 

\end{center}
\vspace{10mm}

\begin{abstract}
  An argument against tolerating finetuning in the Higgs sector is presented,
  by emphasizing the difference between (well understood) quantum corrections
  to scalar masses and the (unsolved) problem of the cosmological constant. I
  also point out that ``split'' supersymmetry, where all scalars except one
  Higgs boson have masses many orders of magnitude above the weak scale, is
  not compatible with simple mechanisms of transmitting supersymmetry breaking
  (gravity, gauge or anomaly mediation), unless a second, independent
  finetuning of parameters is introduced. This finetuning is required to
  obtain an acceptable ratio of vacuum expectation values $\tan\beta$.
\end{abstract}
\clearpage
\setcounter{page}{1}
\pagestyle{plain}

Supersymmetry was originally considered something of a mathematical curiosity
\cite{oldsusy}. It began to be taken seriously as a realistic extension of the
Standard Model (SM) of particle physics only after it was realized
\cite{hierarchy} that it can solve the finetuning problem of the scalar sector
of the SM, by canceling all quadratically divergent corrections to the mass of
the Higgs boson(s) required to break the electroweak gauge symmetry
spontaneously. This is the primary motivation for introducing supersymmetry in
our description of nature.

Of course, supersymmetry has to be broken; no generally accepted mechanism to
achieve this has as yet emerged. Indeed, supersymmetry breaking is widely
perceived to be the ugly side of supersymmetric extensions of the SM, since it
can easily lead to problems with flavor changing neutral currents (FCNC) and
CP violation. However, these problems are just as easily solved if
supersymmetry breaking terms are sufficiently universal and real. In fact, it
was shown \cite{radbreak} that in the minimal supersymmetric extension of the
SM (MSSM), universal supersymmetry breaking terms generated at a high energy
scale allow for radiative $SU(2) \times U(1)_Y \rightarrow U(1)_{\rm em}$
symmetry breaking, thereby finding a dynamical explanation for the negative
squared mass of the Higgs boson that has to be put ``by hand'' into the
Lagrangian of the SM. This mechanism, as well as general finetuning arguments
\cite{nilles}, indicate that superparticle masses should not lie much above
the weak scale, leading to good prospects for their discovery at future
high--energy colliders such as the LHC \cite{lhc}. Moreover, radiative $SU(2)
\times U(1)_Y$ breaking works best for large top mass. Although it can also be
realized with moderate top mass \cite{luis}, the fact that the top quark is
the heaviest known elementary particle can therefore be counted as argument in
favor of softly broken weak--scale supersymmetry \cite{kane}.

A little later it was realized \cite{dm} that broken supersymmetry also
provides a good candidate for the Cold Dark Matter (CDM) which we now believe
to form some 85\% of all matter in the universe, and to contribute about 25\%
of its energy density \cite{wmap}. Finally, measurements at LEP showed
conclusively \cite{amaldi} that in the SM the gauge couplings do not unify at
a point, whereas at the MSSM unification at the percent level can easily be
achieved.\footnote{Exact unification is not expected in the presence of
  unknown high--scale threshold corrections.} Recently it has been pointed out
\cite{split} that these secondary virtues of the MSSM (and similar
supersymmetric models) are shared by models where almost all scalars have very
large masses, possibly even of order of the unification scale. Since the
superpartners of the fermions come in complete representations of $SU(5)$,
they have no impact (to one--loop order) on whether or not the gauge couplings
unify. As long as the gauginos and higgsinos are kept light, unification will
work more or less like in the MSSM. Moreover, the lightest neutralino can
still make a good CDM candidate, so long as it is not Bino--like. Of course,
one Higgs doublet has to be kept light in order to achieve electroweak
symmetry breaking. The authors of ref.\cite{split} coined the phrase ``split
supersymmetry'' for this kind of model.

These models have attracted a fair amount of interest although they make no
pretense of solving the finetuning problem of the SM, i.e. they abandon the
primary virtue of supersymmetry. The argument given in \cite{split}
essentially amounts to the statement that finetuning anyway seems to be
required to solve the cosmological constant problem, so one might as well
allow finetuning also in other sectors of the theory. Indeed, the ``string
landscape'' is supposed to take care of this little thing for us.

However, this argument brushes over the fact that there is an essential
difference between the cosmological constant and the mass of the Higgs boson.
The former is a macroscopic quantity which can be defined only in the
framework of a theory of gravity, presumably General Relativity or an
extension thereof. An understanding of the connection between the vacuum
energy predicted by quantum field theory and the cosmological constant may
therefore only be possible in the framework of a quantum theory of gravity. In
contrast, the mass of the Higgs boson is a microscopic quantity, which should
be computable using the well--known rules of quantum field theory. It
therefore seems premature, to say the least, to interpret our lack of
understanding of the cosmological constant (which is a very serious
theoretical problem indeed!) as {\it carte blanche} for allowing finetuning
anywhere in our description of nature.

Moreover, realistic models with ``split'' supersymmetry are less easy to
construct than their recent popularity suggests. The reason is that one needs
vacuum expectation values (vevs) for the neutral components of {\em both}
Higgs doublets, $v_1 = \langle H_d^0 \rangle$ and $v_2 = \langle H_u^0
\rangle$, in order to give masses to both the bottom and the top quark. If we
want to keep the Yukawa couplings in the perturbative domain, which is
suggested for a perturbative unification of the gauge couplings,
the parameter $\tan \beta \equiv v_2/v_1$ should lie in the range
\be \label{range}
0.5 \lsim \tan\beta \lsim 100. 
\ee
This quantity can be calculated by minimizing the (tree--level) Higgs
potential \cite{pot}, which can be written as
\be \label{pot}
V_{\rm Higgs} = m^2_{H_u} \left| H_u^0 \right|^2 + m^2_{H_d} \left| H_d^0
\right|^2 - ( B \mu H_u^0 H_d^0 + h.c. ) + \frac {g^2+g_Y^2}{8} \left( 
\left| H_u^0 \right|^2 - \left| H_d^0 \right|^2 \right)^2.
\ee
Here $g$ and $g_Y$ are the $SU(2)$ and $U(1)_Y$ gauge couplings, respectively,
and $\mu$ is the mass parameter coupling the two Higgs superfields in the
superpotential. Minimization of (\ref{pot}) gives \cite{pot}
\be \label{tanb}
\sin 2 \beta = \frac {2 B \mu} {m^2_{H_u} + m^2_{H_d}} .
\ee
In models with ``split'' supersymmetry, one has $m^2_{H_u} \sim {\cal
  O}(m_{\rm weak}^2), \ m^2_{H_d} \sim {\cal O}(m^2_{\rm SUSY}) \gg |\mu|^2
\sim {\cal O}(m^2_{\rm weak})$. All other scalars also have masses ${\cal
  O}(m_{\rm SUSY})$, leading to finetuning at the level $(m_{\rm weak}/m_{\rm
  SUSY})^2$.

Recall that in ``split'' supersymmetry, gaugino masses are assumed to be
${\cal O}(m_{\rm weak})$. This could be achieved by a softly broken $R$
symmetry. However, an $R$ symmetry that allows a supersymmetric $\mu$ term
would forbid a nonvanishing $B$, so that $|B| \sim {\cal O}(m_{\rm weak})$,
leading to $\tan\beta \sim {\cal O} ( m^2_{\rm SUSY} / m^2_{\rm weak})$, and
hence 
\be \label{ms0}
m_{\rm SUSY} \lsim 10 \, m_{\rm weak} \ \ \  {\rm if} \ |B| \sim {\cal
  O}(m_{\rm weak}) ,
\ee
well below the scales usually considered in ``split'' supersymmetry.

One thus has to choose the $R-$charges such that they forbid $\mu$ as well as
the gaugino masses, but allow $B \mu$ to be nonzero. One then faces the
challenge to generate $\mu$ (which conserves supersymmetry but breaks this
$R-$symmetry) and gaugino masses (which break both supersymmetry and this
$R-$symmetry) of roughly the same size. Note that the well--known supergravity
solutions \cite{muprob} of the ``$\mu-$problem'' will not work here. Let us
nevertheless assume that $\mu$ and gaugino masses of the appropriate size have
been produced, and investigate the size of $|B \mu|$ predicted by simple
mechanisms of transmitting supersymmetry breaking to the visible sector.

If supersymmetry breaking is transmitted by gravitational--strength
interactions to the visible sector (which automatically happens once
supersymmetry is embedded in a supergravity theory \cite{nilles}), one expects
$|B| \sim {\cal O}(m_{\rm SUSY})$, leading to $\tan\beta \sim {\cal O}(m_{\rm
  SUSY}/m_{\rm weak})$.  This will be compatible with the constraint
(\ref{range}) only if
\be \label{ms1}
m_{\rm SUSY} \lsim 100 \, m_{\rm weak} \ \ \ \ \ [{\rm gravity \ mediation}: \
|B|  \sim {\cal O}(m_{\rm SUSY})].
\ee
This is sufficient to solve all potential problems with FCNC and CP violation,
but leads to a spectrum more reminiscent of so--called ``inverted hierarchy''
or ``more minimal supersymmetry'' models \cite{inv}, rather than what is
typically considered for ``split'' supersymmetry.

How can the bound (\ref{ms1}) be circumvented? First of all, the upper bound
in (\ref{range}) has been obtained by requiring that $m_b = \lambda_b \langle
H_d^0 \rangle$ be sufficiently large for a Yukawa coupling $\lambda_b \lsim
{\cal O}(1)$. Note that the Higgs potential should be minimized at scale
$m_{\rm SUSY}$, which by assumption is exponentially larger than $m_{\rm
  weak}$. However, below $m_{\rm SUSY}$ the $b-$quark mass runs as in the SM,
i.e. only increases by a factor $\lsim 2$ when going down to scale $m_t$. This
effect has already been included by allowing $\tan\beta$ to be as large as
100.

The limit (\ref{ms1}) can thus only be evaded if $|B| \gg m_{\rm SUSY}$. In
fact, this is easily possible in models with gauge mediated supersymmetry
breaking (GMSB) \cite{gmsb}, where the $B \mu$ term can be generated already
at one--loop level, whereas scalar masses are only generated at two--loop
level.\footnote{In fact, this is a problem for GMSB models with the natural
  choice $m_{\rm SUSY} \sim m_{\rm weak}$; somewhat
  complicated constructions are required to suppress the 1--loop contribution
  to $B \mu$ \cite{gmsb}.} This then allows $|B \mu| \sim 100 \, m_{\rm SUSY}$,
leading to
\be \label{ms2}
m_{\rm SUSY} \lsim 10^4 m_{\rm weak} \ \ \ \ \ [{\rm GMSB:} |B\mu| \sim 100 \,
m_{\rm SUSY}].
\ee
This indeed allows a considerable ``splitting'' of the superparticle
spectrum. However, $m_{\rm SUSY}$ would then still have to be many orders
of magnitude below the scale of Grand Unification, and even well below most
other ``intermediate'' scales (e.g. the Peccei--Quinn scale associated with
the spontaneous breaking of a possible $U(1)$ symmetry that can be used to
rotate away the QCD $\theta-$term \cite{pq}). Moreover, gluinos, while
sufficiently long--lived to have detectable decay lengths, would hardly be
``meta--stable'', which is supposed to be a hallmark of models with ``split''
supersymmetry. 

The third simple mechanism to transmit supersymmetry breaking to the visible
sector goes under the name of anomaly mediation \cite{anom}. It naturally
gives (sufficiently) flavor--universal soft scalar masses not to suffer from
FCNC problems. This mechanism always produces gaugino and sfermion masses of
the same order of magnitude; it can therefore not lead to a ``split''
supersymmetry spectrum.

How could even larger values of $|B|$ be generated? Replacing $\mu$ by the vev
of a visible sector singlet field $N$ does not help. In that case $|B| = |A
\langle N \rangle |$, where $A$ is a {\em trilinear} soft breaking parameter,
which contributes to electroweak gaugino masses through finite one--loop
diagrams, and to all gaugino masses (via the $A-$parameters associated with
squarks and sleptons) through two--loop renormalization group equations. The
requirement that gaugino masses are roughly of order of the weak scale would
then give a bound on $m_{\rm SUSY}$ which is stronger than that in
(\ref{ms2}).

The only known way to produce the required $|B| \gg m_{\rm SUSY}$ relies on
$D-$term supersymmetry breaking \cite{giudice} with direct (tree--level)
coupling to the visible sector. Of course, this has been attempted in the
early days of supersymmetry phenomenology \cite{nilles}; it had been largely
abandoned since it is difficult to find non--anomalous models where all
squared sfermion masses are positive. Acceptable models tend to be quite
baroque; for example, the model of ref.\cite{giudice} needs six SM singlets,
plus a tripling of the MSSM matter fields by introducing vector--like partners
for each MSSM matter superfield.\footnote{This turns all gauge beta--functions
  positive. Perturbative unification then requires that these extra matter
  fields should have masses $\gsim 10^{11}$ GeV.} A very recent model
\cite{babu} based on an anomalous $U(1)$ requires an additional strongly
interacting sector (i.e. an additional scale in the theory), and leads to an
only moderately split spectrum, with sfermions not much above the bound
(\ref{ms2}).

In the theoretically well motivated, (comparatively) simple scenarios with
gravity-- or gauge--mediated supersymmetry breaking the bounds (\ref{ms1}),
(\ref{ms2}) can only be evaded if one tunes $m^2_{H_d} \ll m^2_{\rm SUSY}$.
For example, $|B| \sim {\cal O}(m_{\rm SUSY})$ would require $m^2_{H_d} \lsim
{\cal O}( 100\, m_{\rm weak} \, m_{\rm SUSY})$. This would require a
{\em second, independent} finetuning if $m_{\rm SUSY}$ is above the range
(\ref{ms1}), over and above the finetuning required to keep $m^2_{H_u}$ of
order $m^2_{\rm weak}$.

In summary, I have argued that the (unsolved) problem of the cosmological
constant should not be interpreted as evidence in favor of finetuning in other
sectors of the theory. Moreover, a very large splitting between the weak scale
and the scale of sfermion masses can be achieved in simple, well--motivated
models of supersymmetry breaking only if the mass of the second Higgs doublet
is intermediate between these scales, i.e. well below the masses of the
sfermions. Imposing such a hierarchy between the masses of heavy scalars would
require a second, independent finetuning, in addition to that needed to have
$m_{\rm weak} \ll m_{\rm SUSY}$. This should be of concern in any
probabilistic (e.g. ``string landscape'') interpretation of this kind of
model.

\end{document}